\documentclass[showpacs,amsmath,amsart,twocolumn,showkeys,pra,superscriptaddress]{revtex4-1}
\usepackage{dcolumn}    
\usepackage{ifpdf}
\usepackage{amssymb,lineno,amsfonts}
\usepackage{graphicx}   
\usepackage{bm}         
\usepackage{bbm}
\usepackage{mathrsfs}
\usepackage{upgreek}
\usepackage{mathtools}
\usepackage{epstopdf}
\usepackage{setspace}
\usepackage{hyperref}
\usepackage{float}
\usepackage{natbib}
\usepackage[usenames,dvipsnames]{xcolor}
\usepackage[matrix,frame,arrow]{xypic}

\newcommand{\ket}[1]{\vert{#1}\rangle}
\newcommand{\bra}[1]{\langle{#1}\vert}

\newcommand{\inpr}[2]{\langle{#1}\vert{#2}\rangle}
\newcommand{\expec}[1]{\langle{#1}\rangle}
\newcommand{\proj}[1]{\vert{#1}\rangle\langle{#1}\vert}
\definecolor{med-blue}{RGB}{25,25,112}
\hypersetup{colorlinks, linkcolor={red},citecolor={blue}, urlcolor={MidnightBlue}}

\begin{document}

\title{How many runs ensure quantum fidelity in teleportation experiment?}

\author{C S Sudheer Kumar} 
\email{sudheer.kumar@students.iiserpune.ac.in}
\affiliation{NMR Research Center and Department of Physics, \\ Indian Institute of Science Education and Research, Pune 411008, India}
\affiliation{Harish-Chandra Research Institute, HBNI, Chhatnag Road, Jhunsi, Allahabad 211 019, India} 
\author{Ujjwal Sen}
\email{ujjwal@hri.res.in} 
\affiliation{Harish-Chandra Research Institute, HBNI, Chhatnag Road, Jhunsi, Allahabad 211 019, India}

\begin{abstract}
The strong law of large numbers asserts that experimentally obtained mean values, in the limit of number of repetitions of the experiment going to infinity, converges almost surely 
to the theoretical predictions which are based on \textit{a priori} assumed constant values for  probabilities of the random events. Hence in most theoretical 
calculations, we implicitly neglect fluctuations around the mean. However in practice, we can repeat the experiment only finitely many times, and hence fluctuations are inevitable, and 
may lead to erroneous judgments. 
It is theoretically possible to teleport an unknown quantum state, using entanglement, with unit fidelity. The experimentally achieved values are however sub-unit, and often, significantly so.
We show that when the number of repetitions of the experiment is small, there is significant probability of achieving a sub-unit 
experimentally achieved quantum teleportation fidelity that uses entanglement, 
even classically, i.e., without using entanglement. We further show that only when the number of repetitions of the experiment is of the order of a few thousands, the 
probability 
of a 
classical teleportation process to reach the currently achieved experimental quantum teleportation fidelities becomes negligibly small, and hence ensure that the experimentally obtained fidelities are due to genuine use of the shared entanglements.
\end{abstract}
\maketitle

\section{Introduction}
Quantum measurement outcomes being intrinsically random, fluctuations in experimentally obtained mean values are inevitable. However in theoretical considerations of quantum mechanics, we often implicitly neglect fluctuations, e.g., in calculating bounds corresponding to a Bell test \cite{Bell_spek_unspek_book, CHSH_original_paper}, classical limit of teleportation fidelity for quantum states \cite{Teleport_benet_orginal,Nielsen_horodecky-1, Nielsen_horodecky-2}, etc. This is justified by 
Kolmogorov's
strong law of large numbers (LLN) which asserts that 
the experimentally obtained mean values, in the limit of the number of repetitions of the experiment under identical conditions going to infinity, converges almost surely, and under certain assumptions 
(see Appendix \ref{LLN_asumptnsApp} for details),
to the theoretical predictions which are based on \emph{a priori} assumed constant values for the probabilities of the random events \cite{Prob_book_Chow, Prob_book_AlanGut, Sheldon_Ross_probability_book,
probability-Gupta}. It is to be noted that the convergence shown by strong LLN is not pointwise \cite{FQMarXiv}. 
However in practice it is not feasible to repeat the experiment, under identical conditions, infinitely many times. Hence in practice, fluctuations would potentially matter. 

In practical scenarios, as we can repeat the experiment only finitely many times, theoretical  bounds no more apply strictly, because they correspond to an infinite number of repetitions of the experiment. Therefore, for a given finite number of runs of the experiment, it is important to know the probability with which the experimentally obtained mean values of a system characteristic can go beyond the theoretical bounds of the characteristic. In this respect, it is interesting to note that it is possible to achieve the Tsirelson bound \cite{Tsirelsonboundoriginalpaper} of the Bell Clauser-Horne-Shimony-Holt expression, even with unentangled states (via fake correlations arising due to randomness in measurement outcomes), but the probability of achieving it reduces exponentially with the number of repetitions of the experiment, as shown by Gill \cite{Gill_Bellineq_finitesamplsize}. 

Here we consider the problem of teleporting sets of quantum states from one location to another and the theoretical upper bounds for the fidelity of teleporting the quantum states ``classically'' i.e., without using entanglement.
These upper bounds for the different sets of quantum states to be teleported, are based on \textit{a priori} assumed constant values for the probabilities of random events, and hence cannot be observed in actual experiments, due to non-zero fluctuations in the relative frequencies. We show that when the number of repetitions of the experiment is not so large, there is significant probability of the relative frequencies going much beyond the classical teleportation fidelity upper bounds, even if there is no shared entanglement. It is theoretically known that using entanglement, it is possible to teleport an unknown quantum state with unit quantum fidelity \cite{Teleport_benet_orginal}.
We consider data from various experiments which have demonstrated quantum teleportation \cite{Teleport_expt_nature, Teleport_expt_prl_popescu,Teleport_NMR_nature, Teleport_expt_qubitBarnett_nature,Teleport_expt_qubitatomphoton_nature,Tele_expt_qubitYbion, Teleport_expt_qubitMUBnature, Tele_expt_qubitphotontosolid_nature,Teleport_expt_4asymqubt_nature, Tele_expt_NVcentr, Teleport_review_nature, Qutrit_teleport_exptPRL}.
Such experiments typically achieve  sub-unit fidelities.
We provide lower bounds on the number of repetitions of the experiment that is needed to ensure that a sub-unit quantum teleportation fidelity cannot be achieved classically (i.e., without entanglement)
with appreciable probability. 
We find that, to be pretty confident about the teleportation fidelity being genuinely quantum, one needs to typically repeat the 
experiment
a few thousand times.

The rest of the paper is organized as follows. In Sec. \ref{Hoefding_bound}, we formally consider the case of a teleportation experiment 
when the number of repetitions of the experiment is finite, and find the probability of the relative frequency corresponding to classical teleportation going beyond the optimal theoretical classical teleportation fidelity valid in the limit of an infinite number of runs.
In Sec. \ref{ajut-barsha-periye}, 
we apply the result from the preceding section to several cases, by using experimentally obtained 
quantum
teleportation fidelities in the respective cases from the literature. 
In Sec. \ref{sahaj-parinay}, we compare the method pursued in this paper with the method that may use the apparatus of hypothesis testing.
Finally, we conclude in Sec. \ref{conclud}.

\section{Probability of obtaining nonclassical teleportation fidelity values 
without entanglement for finite runs}\label{Hoefding_bound}

Quantum teleportation \cite{Teleport_benet_orginal} is a protocol by which the quantum state of a physical system, that is known to be produced by a source described by the ensemble \(\mathcal{E}=\{p_i, |\psi_i\rangle\}\),  appears as the state of another system, possibly at a distant location. The resources used are (a) local quantum operations at the respective laboratories at the two locations and classical communication between the laboratories, and (b) a shared quantum state. The shared quantum state is of course useful only if it cannot be created by the local operations and classical communication allowed as resource (a), and these are precisely the entangled shared quantum states \cite{eiTa-entanglement-1, eiTa-entanglement-2, eiTa-entanglement-3}.

The transfer of the quantum state in a quantum  teleportation protocol is exact for certain entangled shared states. For others, it only provides a sub-unit average fidelity of the state transfer, that depends both on the shared state and the input ensemble \(\mathcal{E}\). 
Shared entangled states are usually costly to prepare experimentally, and thus it is important to find out whether its inclusion is useful to the protocol. To this end, we consider an average fidelity of quantum state transfer for the case when the shared entangled state is not available as a resource for the state transfer. 
This fidelity is usually referred to as the 
``classical'' teleportation fidelity. It is a function of  the input ensemble \(\mathcal{E}\).
Therefore, a given strategy of quantum teleportation of a given input ensemble by employing local quantum operations and classical communication on a given shared entangled state and the given input ensemble is nonclassical, if the  teleportation fidelity is higher than the corresponding classical fidelity. It is clear that if the ensemble consists of a set of orthogonal states, the classical fidelity is already unity. (See Ref. \cite{puratan} in this regard.) Also, a shared non-entangled state can, by definition, never have a fidelity higher than classical for any teleportation protocol.

The impossibility of an unentangled state to have a teleportation fidelity beyond  the relevant classical teleportation fidelity is however valid 
only in the limit of number of repetitions of the experiment  tending to infinity. But in practice we can repeat an experiment only finitely many times. In this case, there will always be a positive probability of 
a classical teleportation strategy -- a teleportation strategy that does not involve any shared entangled state -- 
to possess a 
teleportation fidelity that is greater than the corresponding theoretical classical teleportation fidelity. 
We are now going to obtain an upper bound for 
this
probability.


\textit{Classical teleportation of unknown quantum states}: Let $X$ be a random variable which takes value $i$ with probability $p_i$, with the latter being a real number between 0 and 1, $i=1,2,...,a$, such that $\sum_{i=1}^{a}p_i=1$. Charu prepares a qudit (a $d-$level quantum system) in the state $\ket{\psi_i}$ if $X=i$ and gives it to Alice, where $\ket{\psi_i}$ is a vector in the $d-$dimensional complex Hilbert space, $i=1,2,...,a$. And $\ket{\psi_i}$'s are in general normalized, but may be nonorthogonal. Alice knows Charu's state preparation procedure but she does not know which one of the states $\ket{\psi_i}$ has been given to her. Alice wants to classically teleport (i.e., without using entanglement) $\ket{\psi_i}$ to Bob. Alice carries out a positive operator valued measurement (POVM) on $\ket{\psi_i}$ with POVM elements $\Pi_l,l=1,2,...,a$ which satisfy the resolution of the identity i.e., $\sum_{l=1}^{a}\Pi_l=\mathbbm{1}_d$. If $\Pi_{l}$ clicks in Alice's measurement, she sends the information about the result of measurement, $l$, to Bob via a classical channel. The classical channel is assumed to be noiseless. Bob, thereafter prepares the state $\ket{\psi_l}$ and sends it to the examiner Debu. Charu has informed Debu about the actual state $\ket{\psi_i}$ which was given to Alice. Debu measures the state sent by Bob, viz. $\ket{\psi_l}$, on the projectors $\{\proj{\psi_i},\mathbbm{1}_d-\proj{\psi_i}\}$ for verification.
Then the theoretical optimal average fidelity of classical teleportation 
is given by 
\begin{eqnarray}
F^{\mathrm{th}}_{\mathrm{cla}}=\max_{\{\Pi_l\}}\sum_{i,l=1}^{a} p_i
\bra{\psi_i}\Pi_l\ket{\psi_i}
|\inpr{\psi_i}{\psi_l}|^2,
\end{eqnarray}
which is really the probability that the state sent by Alice to Bob, passes Debu's test.


 Define $V^{(i)}=\proj{\psi_i}$. Let $V^{(i)}_{\ket{\psi_k}}$ be the outcome of measuring $\{V^{(i)}, \mathbbm{1}_d - V^{(i)}\}$ on the state ${\ket{\psi_k}}$ (so that  $V^{(i)}_{\ket{\psi_k}}$ is a random variable which takes the values $1$ or $0$). Let $\{\Pi_{l}\}_{\ket{\psi_i}}$ be the outcome of the POVM carried out by Alice on the state  $\ket{\psi_i}$ (so that  $\{\Pi_{l}\}_{\ket{\psi_i}}$ is a random variable which outputs one of the following values: $1,2,...,a$). And let $N_y(Y,M)$ be the number of outcomes for which $Y=y$ in $M$ independent trials of the random variable $Y$. Then the experimentally observed 
 fidelity of classical teleportation is given by
\begin{eqnarray}
F^{\mathrm{expt}}_{\mathrm{cla}}=\frac{1}{N}\sum_{i=1}^{a}\bigg(N_i(\{\Pi_{l}\}_{\ket{\psi_i}},N_i(X,N))\nonumber\\+\sum_{k\ne i;k=1}^{a}N_1\big(V^{(i)}_{\ket{\psi_k}},N_k(\{\Pi_{l}\}_{\ket{\psi_i}},N_i(X,N))\big)\bigg),
\label{Fexpt}
\end{eqnarray}
which is subject to the constraints 
\begin{eqnarray}
\sum_{k=1}^{a}N_k(\{\Pi_{l}\}_{\ket{\psi_i}},N_i(X,N))=N_i(X,N)\nonumber\\
\mathrm{and}~\sum_{i=1}^{a}N_i(X,N)=N.
\end{eqnarray}
Let $P(\cdots)$ represent the probability of the event in its argument. The strong LLN asserts that \cite{Sheldon_Ross_probability_book}
\begin{eqnarray}
P\bigg(\lim\limits_{N\rightarrow\infty}F^{\mathrm{expt}}_{\mathrm{cla}}=F^{\mathrm{th}}_{\mathrm{cla}}\bigg)=1.
\label{LLN}
\end{eqnarray}
(See 
Appendix \ref{LLNderiveapp} for a derivation.) One should note that the convergence implied by the strong LLN is not pointwise. 
Eq. (\ref{LLN}) implies that the event $\lim_{N\rightarrow\infty}F^{\mathrm{expt}}_{\mathrm{cla}}=F^{\mathrm{th}}_{\mathrm{cla}}$ happens almost surely but not 100\% surely \cite{Prob_book_weigh_odd, circularLLNSpanos2013,Prob_book_AlanGut}. This is because there is a nonempty  measure-zero set (which is infinitely large) which corresponds to the event $\lim_{N\rightarrow\infty}F^{\mathrm{expt}}_{\mathrm{cla}}=F^{\mathrm{th}}_{\mathrm{cla}}$ not happening \cite{Prob_book_weigh_odd, circularLLNSpanos2013}. Moreover, in any real experiment, $N<\infty$, and hence we have the possibility that $P(F^{\mathrm{expt}}_{\mathrm{cla}}\ge F^{\mathrm{th}}_{\mathrm{cla}}+t)>0$ where $t>0$. We are now going to find an upper bound for $P(F^{\mathrm{expt}}_{\mathrm{cla}}\ge F^{\mathrm{th}}_{\mathrm{cla}}+t)$, corresponding to a given $N$. Gill has obtained similar upper bound, in the context of a Bell inequality \cite{Gill_Bellineq_finitesamplsize}.

An experimentalist (call her Dodolu) performing a quantum teleportation experiment has 
the
intention of convincing us that $F^{\mathrm{th}}_{\mathrm{qm}}>F^{\mathrm{th}}_{\mathrm{cla}}$, where $F^{\mathrm{th}}_{\mathrm{qm}}$ is the theoretically calculated quantum mechanical teleportation fidelity. $F^{\mathrm{th}}_{\mathrm{qm}}$ is always unity, assuming that a maximally entangled state is used for the teleportation. However, Dodolu 
only gets $F^{\mathrm{expt}}_{\mathrm{qm}}$, which is the experimentally obtained  quantum mechanical teleportation fidelity, and she is forced to replace $F^{\mathrm{th}}_{\mathrm{qm}}$ in all relations by $F^{\mathrm{expt}}_{\mathrm{qm}}$. Hence, her intention now is to demonstrate that $F^{\mathrm{expt}}_{\mathrm{qm}}>F^{\mathrm{th}}_{\mathrm{cla}}$.

One must now be careful to make sure that had somebody (call him Eneet) performed an experiment without entanglement, then $F^{\mathrm{expt}}_{\mathrm{cla}}$ does not reach $F^{\mathrm{expt}}_{\mathrm{qm}}$ easily, due to the finite number of runs performed in the experiment. It may be noted that the previous experiment (of Dodolu) becomes the latter one (of Eneet) in the absence of entanglement.

Let us make the following assumptions to simplify the calculations:
\begin{itemize}
\item 
Let $p_i=1/a~\forall i$. 
\item 
Let us neglect, assuming $N$ to be sufficiently large, the fluctuation of $N_i(X,N)/N$ around its theoretical mean value, i.e., $1/a$. Then $N_i(X,N)/N\approx 1/a~\forall i$. 
\end{itemize}
 Note that for a given $N$, the upper bound of $P(F^{\mathrm{expt}}_{\mathrm{cla}}\ge F^{\mathrm{th}}_{\mathrm{cla}}+t)$ considering the fluctuation of $N_i(X,N)/N$, will be higher than the upper bound of $P(F^{\mathrm{expt}}_{\mathrm{cla}}\ge F^{\mathrm{th}}_{\mathrm{cla}}+t)$ without considering the fluctuation of $N_i(X,N)/N$. Using the constraint 
 \begin{equation*}
 N_i(\{\Pi_{l}\}_{\ket{\psi_i}},N/a)=N/a-\sum_{k\ne i;k=1}^{a}N_k(\{\Pi_{l}\}_{\ket{\psi_i}},N/a),
 \end{equation*}
 we eliminate $N_i(\{\Pi_{l}\}_{\ket{\psi_i}},N/a)$ from Eq. (\ref{Fexpt}), to obtain
\begin{eqnarray}
F^{\mathrm{expt}}_{\mathrm{cla}}=\frac{1}{N}\sum_{i=1}^{a}\bigg(N/a-\sum_{k\ne i;k=1}^{a}N_k(\{\Pi_{l}\}_{\ket{\psi_i}},N/a)\nonumber\\+\sum_{k\ne i;k=1}^{a}N_1(V^{(i)}_{\ket{\psi_k}},N_k(\{\Pi_{l}\}_{\ket{\psi_i}},N/a))\bigg),
\label{Fexpt1}
\end{eqnarray}
where we have assumed $N$ to be an integer multiple of $a$. Let $N_{y,z}((Y,Z),M)$ be the number of outcomes in which $Y=y$ and $Z=z$ in $M$ independent trials each of $Y$ and $Z$. Then using the identity 
\begin{equation*}
N_{y,z}((Y,Z),M)=N_{z}(Z,N_y(Y,M)),
\end{equation*}
 Eq. (\ref{Fexpt1}) reduces to
\begin{eqnarray}
F^{\mathrm{expt}}_{\mathrm{cla}}=1+\frac{1}{N}\sum_{i=1}^{a}\sum_{k\ne i;k=1}^{a}\sum_{n=1}^{N/a}\tilde{\Pi}^{(k)}_{\ket{\psi_i},n}(V^{(i)}_{\ket{\psi_k},n}-1),\quad
\label{Fexpt2}
\end{eqnarray}
where $\tilde{\Pi}^{(k)}_{\ket{\psi_i},n}=1$ if the outcome in the $n^{\mathrm{th}}$ trial of $\{\Pi_{l}\}_{\ket{\psi_i}}$ is $k$, and $\tilde{\Pi}^{(k)}_{\ket{\psi_i},n}=0$ otherwise; and $V^{(i)}_{\ket{\psi_k},n}$ is the outcome in the $n^{\mathrm{th}}$ trial of $V^{(i)}_{\ket{\psi_k}}$. Note that in all further calculations we are going to use $F^{\mathrm{expt}}_{\mathrm{cla}}$ as given in Eq. (\ref{Fexpt2}), and 
\begin{equation*}
F^{\mathrm{th}}_{\mathrm{cla}}=(1/a)\sum_{i,l=1}^{a}\bra{\psi_i}\Pi_l\ket{\psi_i}|\inpr{\psi_i}{\psi_l}|^2.
\end{equation*}

We now state a result that will be useful for our purposes. Let $Y_1,Y_2,...,Y_m$ be independent random variables with finite first and second moments, and
\begin{eqnarray}
S_m=Y_1+Y_2+...+Y_m,~\bar{Y}=\frac{S_m}{m},~\mu=\expec{\bar{Y}}=\frac{\expec{S_m}}{m}.\nonumber
\end{eqnarray}
\emph{Hoeffding bound}: Let $t'=t/(\tilde{b}-\tilde{a}),t>0$ and $\mu'=(\mu-\tilde{a})/(\tilde{b}-\tilde{a})$. If $Y_1,Y_2,...,Y_m$ are independent random variables with $\tilde{a}\le Y_j\le \tilde{b}$ for $j=1,2,...,m$ then for $0<t'<1-\mu'$,
\begin{eqnarray}
P(\bar{Y}\ge \mu+t)\le \bigg(\big(\frac{\mu'}{\mu'+t'}\big)^{\mu'+t'}\big(\frac{1-\mu'}{1-\mu'-t'}\big)^{1-\mu'-t'}\bigg)^m.\nonumber\\ 
\end{eqnarray}
For the proof of the preceding bound, see \cite{Hoefding_prob_ineq_finitesamplsize}. For a refined Hoeffding bound, see \cite{HoefdingBoundImprovd} (see also \cite{HoefdingBoundImprcd1}). We note that the bound is exact (i.e., we have not used any approximations like Gaussian, Poisson, etc. distributions for the actual distribution) but not tight (i.e., not optimal). However it is better than the earlier bounds available in the literature (see \cite{Hoefding_prob_ineq_finitesamplsize} in this regard).

In the Hoeffding bound, let us set 
\begin{equation*}
Y_j=\tilde{\Pi}^{(k)}_{\ket{\psi_i},n}(V^{(i)}_{\ket{\psi_k},n}-1),
\end{equation*}
where $j=ikn$, and $1\le j\le m$. This implies that
\begin{eqnarray}
 S_m=\sum_{i=1}^{a}\sum_{k\ne i;k=1}^{a}\sum_{n=1}^{N/a}\tilde{\Pi}^{(k)}_{\ket{\psi_i},n}(V^{(i)}_{\ket{\psi_k},n}-1).\nonumber
\end{eqnarray}
Here, $m=(a-1)N,a>1$, $\tilde{a}=-1,\tilde{b}=0$, $F^{\mathrm{expt}}_{\mathrm{cla}}=1+(a-1)\bar{Y}$, and
\begin{equation}
\expec{F^{\mathrm{expt}}_{\mathrm{cla}}}=1+(a-1)\mu=F^{\mathrm{th}}_{\mathrm{cla}}.
\label{Fthcla}
\end{equation}
 Then, the Hoeffding bound simplifies to
\begin{eqnarray}
&&P(F^{\mathrm{expt}}_{\mathrm{cla}}\ge F^{\mathrm{th}}_{\mathrm{cla}}+(a-1)t)\nonumber\\
&&\le\bigg(\big(\frac{\mu+1}{\mu+1+t}\big)^{\mu+1+t}\big(\frac{\mu+t}{\mu}\big)^{\mu+t}\bigg)^{(a-1)N}.
\label{PFexptcla_mainresult}
\end{eqnarray}

\section{When is teleportation quantum?}
\label{ajut-barsha-periye}
We now consider experimentally obtained fidelities of quantum teleportation from the literature, in various 
paradigmatic situations.
These experiments utilize shared entangled states for realizing the corresponding fidelities. Then for a given number, $N$, of runs of the experiment, we calculate the upper bound for the probability of achieving the same fidelity experimentally but via classical teleportation. This will tell us how many repetitions of the experiment are necessary to make sure that the experimentally achieved teleportation fidelity is genuinely quantum.
\subsection{Teleporting three symmetric qubit states}\label{qubittele3}
We begin by considering quantum teleportation of the three qubit states given by \cite{Teleport_expt_prl_popescu, stat_discrm_barnet_review}
\begin{eqnarray}
\ket{\psi_1}=\ket{0},~\ket{\psi_2}=(\ket{0}-\sqrt{3}\ket{1})/2,\nonumber\\
\ket{\psi_3}=(\ket{0}+\sqrt{3}\ket{1})/2,
\end{eqnarray}
where $\ket{0}$ and $\ket{1}$ are the eigenstates of Pauli-z matrix $\sigma_z$ with eigenvalues $+1$ and $-1$ respectively. These three states are placed symmetrically on a great circle of the Bloch sphere. An experimentally obtained teleportation fidelity is
$F^{\mathrm{expt}}_{\mathrm{qm}}=0.865$ \cite{Teleport_expt_prl_popescu}. We show that at small $N$, there is significant probability of experimentally achieving, even via classical teleportation (i.e., without using entanglement), the same amount of fidelity (i.e., 0.865), otherwise achievable only via quantum teleportation. This shows that we should take $N$ sufficiently large enough to make sure that the high fidelity we achieve is genuinely quantum in origin, i.e., due to using entanglement.


The square-root measurement is defined in \cite{stat_discrm_barnet_review} as follows: $\Pi_l=(2/3)\proj{\psi_l},~l=1,2,3$. Now, ``for many of
the cases in which the optimal minimum error measurement is known, it is the
square-root measurement'' \cite{stat_discrm_barnet_review}. One can verify that $\sum_{l=1}^{3}\Pi_l=\mathbbm{1}_2$. Then we obtain $F^{\mathrm{th}}_{\mathrm{cla}}=3/4$ (which is same as the classical upper bound derived in \cite{Teleport_expt_prl_popescu}), $\mu=-1/8$, and $t=0.0575$. Note that the value of $\mu$ is dictated by the set $\{\ket{\psi_i}\}$ and the optimal measurement strategy to perform minimal error discrimination on this set, and by the relations $\mu=\expec{S_m}/m$ and Eq. (\ref{Fthcla}). On the other hand, the value of $t$ is obtained by equating $F^{\mathrm{th}}_{\mathrm{cla}}+(a-1)t$ with the experimentally obtained value $F^{\mathrm{expt}}_{\mathrm{qm}}$. Substituting these values in Eq. (\ref{PFexptcla_mainresult}), we obtain the following upper bounds:
\begin{eqnarray}
\mathrm{for}~ N&=&100,\phantom{0} P(F^{\mathrm{expt}}_{\mathrm{cla}}\ge 0.865)\le 0.0293,\nonumber\\
  \mathrm{for}~ N&=&1000, P(F^{\mathrm{expt}}_{\mathrm{cla}}\ge 0.865)\le 4.625\times 10^{-16},\nonumber\\
   \mathrm{for}~ N&=&5000, P(F^{\mathrm{expt}}_{\mathrm{cla}}\ge 0.865)\le 2.116\times 10^{-77}.\nonumber\\
\end{eqnarray}
Note that the difference in the bound corresponding to $N=99$ (an integer multiple of $a=3$) and $N=100$ is insignificant. Moreover, by the principle of interpolation, the bound given in expression (\ref{PFexptcla_mainresult}) holds to a good extent, for $N$ which is not an integer multiple of $a$ as well.

If it would have been possible to experimentally obtain a quantum fidelity very close to unity, say \(1-10^{-5}\),
then this analysis would have been of little interest. This is because, in that case, 
\begin{eqnarray}
\mathrm{for}~N=50,\phantom{0}~P(F^{\mathrm{expt}}_{\mathrm{cla}}\ge 1-10^{-5})\le1.597\times 10^{-6},\nonumber\\
\mathrm{for}~N=100,~P(F^{\mathrm{expt}}_{\mathrm{cla}}\ge 1-10^{-5})\le 2.55\times 10^{-12},\nonumber\\
\mathrm{for}~N=500,~P(F^{\mathrm{expt}}_{\mathrm{cla}}\ge 1-10^{-5})\le1.08\times 10^{-58},\nonumber\\
\end{eqnarray}
so that
even for small $N$, the probabilities are negligibly small.

\subsection{Teleportation of four asymmetric qubit states}\label{qubittele4asym}
In this subsection, we consider 
quantum teleportation of the following four  qubit states, that, unlike the case in the preceding subsection, are asymmetrically placed on the Bloch sphere \cite{Teleport_expt_4asymqubt_nature}:
\begin{eqnarray}
\ket{\psi_1}=\ket{0},~\ket{\psi_2}&=&\ket{1},~\ket{\psi_3}=(\ket{0}-i\ket{1})/\sqrt{2},\nonumber\\
\mathrm{and}~\ket{\psi_4}&=&(2\ket{0}-\ket{1})/\sqrt{5}.\nonumber
\end{eqnarray}
An experimentally obtained teleportation fidelity is $F^{\mathrm{expt}}_{\mathrm{qm}}=0.875$ \cite{Teleport_expt_4asymqubt_nature}.
Let $\rho=\frac{1}{4}\sum_{i=1}^{4}\proj{\psi_i}$. The square-root measurement is given in the current scenario as follows
\cite{stat_discrm_barnet_review}:
\begin{eqnarray}
\Pi_l=\frac{1}{4}\rho^{-1/2}\proj{\psi_l}\rho^{-1/2},~l=1,2,3,4.
\end{eqnarray}
One can verify that $\sum_{l=1}^{4}\Pi_l=\mathbbm{1}_2$. Then we obtain $F^{\mathrm{th}}_{\mathrm{cla}}=0.777$, 
$\mu=-0.0743$, and $t=0.0327$. Substituting these values in Eq. (\ref{PFexptcla_mainresult}), we obtain the following upper bounds:
\begin{eqnarray}
 \mathrm{for}~ N&=&100,\phantom{0} P(F^{\mathrm{expt}}_{\mathrm{cla}}\ge 0.875)\le0.0649,\nonumber\\
  \mathrm{for}~ N&=&1000, P(F^{\mathrm{expt}}_{\mathrm{cla}}\ge 0.875)\le 1.3219\times 10^{-12},\nonumber\\
 \mathrm{for}~  N&=&5000, P(F^{\mathrm{expt}}_{\mathrm{cla}}\ge 0.875)\le 4.0362\times 10^{-60}.\nonumber\\
\end{eqnarray}

\subsection{Teleporting qubit MUBs}\label{qubitmubtele}
Here
we consider 
quantum teleportation of three maximally unbiased bases (MUBs) of a qubit.
When the Hilbert space dimension $d$ is a prime power, 
there exist
sets of $d+1$ MUBs. These sets are maximal in the sense
that it is not possible to find more than $d+1$ MUBs in any $d$-dimensional Hilbert
space \cite{MUBWOOTTERS1989363, MUBBandyopadhyay2002, MUB_review_durt, MUB_Wooters1}. They are maximally unbiased in the sense that the absolute value of the inner product between any vector in any of the bases with any vector in any other basis is \(1/\sqrt{d}\). The elements of any basis are mutually orthonormal.

Let $\ket{\pm}=(\ket{0}\pm\ket{1})/\sqrt{2}$, $\ket{\pm y}=(\ket{0}\pm i\ket{1})/\sqrt{2}$. 
Then 
\begin{eqnarray}
\{\ket{0},\ket{1}\},~\{\ket{+},\ket{-}\}, ~\mathrm{and}~\{\ket{+y},\ket{-y}\}
\end{eqnarray}
form a maximal set of MUBs of a qubit.
An experimentally obtained teleportation fidelity is $F^{\mathrm{expt}}_{\mathrm{qm}}=0.77$ \cite{Teleport_expt_qubitMUBnature,Tele_expt_NVcentr}.

Let 
\begin{eqnarray}
\ket{\psi_1}&=&\ket{0},~\ket{\psi_2}=\ket{1},~\ket{\psi_3}=\ket{+},~\ket{\psi_4}=\ket{-}\nonumber\\
\ket{\psi_5}&=&\ket{+y},~\ket{\psi_6}=\ket{-y},\nonumber
\end{eqnarray}
and $\rho=\frac{1}{6}\sum_{i=1}^{6}\proj{\psi_i}=\mathbbm{1}_2/2$. Then the corresponding square-root measurement is given by 
\cite{stat_discrm_barnet_review}:
\begin{eqnarray}
\Pi_l=\frac{1}{6}\rho^{-1/2}\proj{\psi_l}\rho^{-1/2},~l=1,2,\ldots,6.
\end{eqnarray}
One can verify that $\sum_{l=1}^{6}\Pi_l=\mathbbm{1}_2$. We obtain $F^{\mathrm{th}}_{\mathrm{cla}}=2/3$ (which is the same as that obtained by averaging over the entire Bloch sphere \cite{Nielsen_horodecky-1}), $\mu=-1/15$, and $t=0.0207$. Substituting these values in Eq. (\ref{PFexptcla_mainresult}), we obtain the following upper bounds:
\begin{eqnarray}
\mathrm{for}~ N&=&100,\phantom{0} P(F^{\mathrm{expt}}_{\mathrm{cla}}\ge 0.77)\le 0.1468, \nonumber\\
  \mathrm{for}~ N&=&1000, P(F^{\mathrm{expt}}_{\mathrm{cla}}\ge 0.77)\le 4.6336\times 10^{-9}, \nonumber\\
   \mathrm{for}~ N&=&5000, P(F^{\mathrm{expt}}_{\mathrm{cla}}\ge 0.77)\le 2.1359\times 10^{-42}.\nonumber\\
\end{eqnarray}



\subsection{Teleporting basis vectors of qutrit MUBs}\label{qutrittele}
We now consider 
quantum teleportation of the states of the four qutrit MUBs.
 Let
\begin{eqnarray}
\ket{\psi_1}&=&\ket{0},~\ket{\psi_2}=\ket{1},~\ket{\psi_3}=\ket{2},\nonumber\\
\ket{\psi_4}&=&\frac{1}{\sqrt{3}}(\ket{0}+\ket{1}+\ket{2}),\nonumber\\
\ket{\psi_5}&=&\frac{1}{\sqrt{3}}(\ket{0}+\omega\ket{1}+\omega^2\ket{2}),\nonumber\\
\ket{\psi_6}&=&\frac{1}{\sqrt{3}}(\ket{0}+\omega^2\ket{1}+\omega\ket{2}),\nonumber\\
\ket{\psi_7}&=&\frac{1}{\sqrt{3}}(\omega\ket{0}+\ket{1}+\ket{2}),\nonumber\\
\ket{\psi_8}&=&\frac{1}{\sqrt{3}}(\ket{0}+\omega\ket{1}+\ket{2}),\nonumber\\
\ket{\psi_9}&=&\frac{1}{\sqrt{3}}(\ket{0}+\ket{1}+\omega\ket{2}),\nonumber\\
\ket{\psi_{10}}&=&\frac{1}{\sqrt{3}}(\omega^2\ket{0}+\ket{1}+\ket{2}),\nonumber\\
\ket{\psi_{11}}&=&\frac{1}{\sqrt{3}}(\ket{0}+\omega^2\ket{1}+\ket{2}),\nonumber\\
\ket{\psi_{12}}&=&\frac{1}{\sqrt{3}}(\ket{0}+\ket{1}+\omega^2\ket{2}),~~~~~~~~
\end{eqnarray}
where $\ket{0},\ket{1},$ and $\ket{2}$ are mutually orthonormal vectors, 
and $\omega=e^{i2\pi/3}$. For a qutrit, there are four MUBs, and one such set has the following as its elements:
\begin{eqnarray}
\{\ket{\psi_1},\ket{\psi_2},\ket{\psi_3}\},~ \{\ket{\psi_4},\ket{\psi_5},\ket{\psi_6}\},\nonumber\\
 \{\ket{\psi_7},\ket{\psi_8},\ket{\psi_9}\}, ~\mathrm{and}~ \{\ket{\psi_{10}},\ket{\psi_{11}},\ket{\psi_{12}}\}.
\end{eqnarray}
An experimentally obtained teleportation fidelity for transferring these 12 quantum states is $F^{\mathrm{expt}}_{\mathrm{qm}}=0.751$ \cite{Qutrit_teleport_exptPRL}.

Let $\rho=\frac{1}{12}\sum_{i=1}^{12}\proj{\psi_i}=\mathbbm{1}_3/3$. The square-root measurement in this scenario is given by 
\cite{stat_discrm_barnet_review}:
\begin{eqnarray}
\Pi_l=\frac{1}{12}\rho^{-1/2}\proj{\psi_l}\rho^{-1/2},~l=1,2,...,12.
\end{eqnarray}
 One can verify that $\sum_{l=1}^{12}\Pi_l=\mathbbm{1}_3$. We obtain $F^{\mathrm{th}}_{\mathrm{cla}}=1/2$. 
 We note that this is the same as the fidelity of teleportation 
 obtained 
 by averaging over all possible qutrit states. Here we achieve this even without averaging over all qutrit states. This is because we are considering a maximal set of MUBs which is sufficient to tomograph an arbitrary unknown qutrit state \cite{MUB_review_durt}. 
 We also have $\mu=-1/22$ and $t=0.0228$. Substituting these values in Eq. (\ref{PFexptcla_mainresult}), we obtain the following upper bounds: 
\begin{eqnarray}
\mathrm{for}~ N&=&50,\phantom{00} P(F^{\mathrm{expt}}_{\mathrm{cla}}\ge 0.751)\le 0.018, \nonumber\\
\mathrm{for}~ N&=&100,\phantom{0} P(F^{\mathrm{expt}}_{\mathrm{cla}}\ge 0.751)\le 3.228\times 10^{-4},\nonumber\\
 \mathrm{for}~ N&=&1000, P(F^{\mathrm{expt}}_{\mathrm{cla}}\ge 0.751)\le 1.2284\times 10^{-35}. \nonumber\\
\end{eqnarray}

\subsection{Violating the Helstrom bound for finite runs}\label{qubittelehelstrom}
The Helstrom bound was an early and remains a very useful method for discrimination between two arbitrary pure states \cite{Helstrom1, Helstrom2, Helstrom3}. See also \cite{stat_discrm_barnet_review, Helstrom_bound_violat}.

Consider the following two pure qubit states:
\begin{eqnarray}
\ket{\psi_1}=\ket{0},~\ket{\psi_2}=\cos(\theta/2)\ket{0}+\sin(\theta/2)\ket{1}.\nonumber
\end{eqnarray}
Also, set
\begin{eqnarray}
\ket{\phi_1}=\cos\frac{\pi-\theta}{4}\ket{0}-\sin\frac{\pi-\theta}{4}\ket{1},\nonumber\\
\ket{\phi_2}=\sin\frac{\pi-\theta}{4}\ket{0}+\cos\frac{\pi-\theta}{4}\ket{1}.\nonumber
\end{eqnarray}
Consider the measurement operators $\Pi_l=\proj{\phi_l},~l=1,2$. One can verify that $\sum_{l=1}^{2}\Pi_l=\mathbbm{1}_2$. Then the probability of error in discriminating between the $\ket{\psi_i}$'s by measuring $\Pi_l$'s turns out to be
\begin{eqnarray}
P_{\mathrm{err}}=\frac{1}{2}\bra{\psi_1}\Pi_2\ket{\psi_1}+\frac{1}{2}\bra{\psi_2}\Pi_1\ket{\psi_2}\nonumber\\
=\frac{1}{2}(1-\sqrt{1-|\inpr{\psi_1}{\psi_2}|^2})=\frac{1}{2}(1-\sin(\theta/2)),
\end{eqnarray}
which is nothing but the Helstrom bound. 
Then we obtain 
\begin{eqnarray}
F^{\mathrm{th}}_{\mathrm{cla}} 
=1-\sin^2(\theta/2)(1-\sin(\theta/2))/2,
\end{eqnarray}
in the classical teleportation protocol where Alice performs a measurement onto the operators \(\{\Pi_1, \Pi_2\}\) to distinguish between the states \(|\psi_1\rangle\) and \(|\psi_2\rangle\), and accordingly sends the information to Bob.
Now, $\mu=F^{\mathrm{th}}_{\mathrm{cla}}-1$ (from Eq. (\ref{Fthcla}), since $a=2$ here). Let $\theta=\pi/2$. Then $F^{\mathrm{th}}_{\mathrm{cla}}=0.9268$ (this is same as that given in \cite{Teleport_twostat_thery}).
We did not find a value of \(F^{\mathrm{expt}}_{\mathrm{qm}}\) in the literature, and we arbitrarily set it to 0.98. $\Rightarrow t=0.0532$.
Substituting these values in Eq. (\ref{PFexptcla_mainresult}), we obtain the following upper bounds:
\begin{eqnarray}
  \mathrm{for}~ N&=&100,\phantom{0} P(F^{\mathrm{expt}}_{\mathrm{cla}}\ge 0.98 )\le 0.0564 ,\nonumber\\
   \mathrm{for} ~N&=&1000, P(F^{\mathrm{expt}}_{\mathrm{cla}}\ge 0.98)\le 3.2687\times 10^{-13},\nonumber\\
    \mathrm{for}~ N&=&5000, P(F^{\mathrm{expt}}_{\mathrm{cla}}\ge 0.98)\le 3.7313\times 10^{-63}. \nonumber\\
 \end{eqnarray}
 
\section{Connection with hypothesis testing}
\label{sahaj-parinay}
The techniques that we have presented above, to test for possible violation of the classical teleportation fidelity bound by an unentangled state in the physically relevant situation of a finite number of runs of the experiment, is similar in spirit to hypothesis testing in statistics, wherein, depending on the experimentally obtained mean value, a primary hypothesis is either accepted or rejected against an alternative secondary hypothesis \cite{Statistics_hypothesistest_pvalueBook}. In our case, the primary hypothesis is the quantum teleportation fidelity, and secondary hypothesis is the classical teleportation fidelity.

Let us assume that a black box is performing teleportation, and we want to know if it is doing classical (without using entanglement) or quantum (using entanglement) teleportation. Let us assume that the experimentally obtained fidelity $F^{\mathrm{expt}}$ is a normally distributed random variable with variance $\sigma^2/N$, where $N$ is the number of repetitions of the teleportation protocol, and let its mean be either $F^{\mathrm{th}}_{\mathrm{cla}}$ (if the black box is doing classical teleportation) or  $F^{\mathrm{th}}_{\mathrm{qm}}$ (if the black box is doing quantum teleportation). For the sake of illustration, let us further assume $F^{\mathrm{th}}_{\mathrm{qm}}<1$. Let us choose $F^{\mathrm{th}}_{\mathrm{qm}}(>F^{\mathrm{th}}_{\mathrm{cla}})$ to be the primary hypothesis, and $F^{\mathrm{th}}_{\mathrm{cla}}$ to be the secondary (alternate) hypothesis. Let us select a critical value $F^{\mathrm{th}}_{\mathrm{cla}}<F^{\mathrm{expt}}_{\mathrm{c}}<F^{\mathrm{th}}_{\mathrm{qm}}$ for the decision criterion. If $F^{\mathrm{expt}}<F^{\mathrm{expt}}_{\mathrm{c}}$, we will reject the hypothesis, and if $F^{\mathrm{expt}}>F^{\mathrm{expt}}_{\mathrm{c}}$, we will accept the hypothesis. If the primary hypothesis is actually correct but $F^{\mathrm{expt}}<F^{\mathrm{expt}}_{\mathrm{c}}$, then we will be erroneously rejecting primary hypothesis and doing type I error. The probability of such an error is 
\begin{eqnarray}
\alpha&=&P(F^{\mathrm{expt}}<F^{\mathrm{expt}}_{\mathrm{c}})\nonumber\\
&=&\frac{\sqrt{N}}{\sqrt{2\pi}\sigma}\int_{-\infty}^{F^{\mathrm{expt}}_{\mathrm{c}}}\exp(-\frac{(F^{\mathrm{expt}}-F^{\mathrm{th}}_{\mathrm{qm}})^2N}{2\sigma^2})dF^{\mathrm{expt}}.\quad \quad
\end{eqnarray}
If the primary hypothesis is actually wrong but $F^{\mathrm{expt}}>F^{\mathrm{expt}}_{\mathrm{c}}$, then we will be erroneously accepting primary hypothesis and doing type II error. The probability of such an error is 
\begin{eqnarray}
\beta&=&P(F^{\mathrm{expt}}>F^{\mathrm{expt}}_{\mathrm{c}})\nonumber\\
&=&\frac{\sqrt{N}}{\sqrt{2\pi}\sigma}\int_{F^{\mathrm{expt}}_{\mathrm{c}}}^\infty\exp(-\frac{(F^{\mathrm{expt}}-F^{\mathrm{th}}_{\mathrm{cla}})^2N}{2\sigma^2})dF^{\mathrm{expt}}.\quad
\end{eqnarray}
$\alpha$ is known as significance level of the test, and $1-\beta$ is known as the power of the test \cite{Statistics_hypothesistest_pvalueBook}. 
This implies that the
larger the $N$ and/or $F^{\mathrm{th}}_{\mathrm{qm}}-F^{\mathrm{th}}_{\mathrm{cla}}$, the lesser will be the $\alpha,\beta$. This conclusion is the same as our analysis based on the Hoeffding bound. But it is important to note that hypothesis testing uses the normal approximation, whereas the Hoeffding bound is exact. In this sense, the analysis based on the Hoeffding bound is more reliable 
(uses less assumptions)
than hypothesis testing.

\section{Conclusion}\label{conclud}
 We considered various experimentally obtained quantum teleportation fidelities and showed that when the number of repetitions of the experiment is small, there is significant probability of experimentally obtaining the same even classically (i.e., without using entanglement). We further showed that if the number of repetitions of the experiment is sufficiently large - and sometimes even if it is only moderately large, then the probability of experimentally achieving the same classically becomes negligibly small and ensures nonclassical fidelity of teleportation with high probability.
 


\appendix
\section{Strong LLN: Assumptions}\label{LLN_asumptnsApp}
The strong LLN is based on the assumption of existence of a probability space or probability triple $(\Omega,\mathcal{F},P)$ where $\Omega$ is the sample space, $\mathcal{F}$ is the collection of random events, and $P$ is a probability measure \cite{Prob_book_AlanGut}. 
Note that the justification for the 
\emph{a priori} assumption of existence of a probability measure
via strong LLN is circular \cite{Prob_book_Chow} because the convergence shown by strong LLN is not pointwise but in terms of the very notion of probability which it is actually trying to justify. There is another assumption involved which is mentioned in the succeeding appendix. 

\section{Deriving Eq. (\ref{LLN})}\label{LLNderiveapp}
Consider Eq. (\ref{Fexpt}).
\begin{eqnarray}
F^{\mathrm{expt}}_{\mathrm{cla}}=\sum_{i=1}^{a}\bigg(\frac{N_i(\{\Pi_{l}\}_{\ket{\psi_i}},N_i(X,N))}{N_i(X,N)}\frac{N_i(X,N)}{N}\nonumber\\
+\sum_{k\ne i;k=1}^{a}\frac{N_1(V^{(i)}_{\ket{\psi_k}},N_k(\{\Pi_{l}\}_{\ket{\psi_i}},N_i(X,N)))}{N_k(\{\Pi_{l}\}_{\ket{\psi_i}},N_i(X,N))}\nonumber\\
\times\frac{N_k(\{\Pi_{l}\}_{\ket{\psi_i}},N_i(X,N))}{N_i(X,N)}\frac{N_i(X,N)}{N}\bigg).\quad
\label{Fexptapp}
\end{eqnarray}
Assuming $N_i(X,N),N_k(\{\Pi_{l}\}_{\ket{\psi_i}},N_i(X,N))\rightarrow\infty$ as $N\rightarrow\infty$, we obtain using strong LLN,
\begin{eqnarray}
P\bigg(\lim_{N\rightarrow\infty}\frac{N_i(X,N)}{N}=p_i\bigg)=1,\nonumber\\
P\bigg(\lim_{N\rightarrow\infty}\frac{N_k(\{\Pi_{l}\}_{\ket{\psi_i}},N_i(X,N))}{N_i(X,N)}=\bra{\psi_i}\Pi_k\ket{\psi_i}\bigg)=1,\nonumber\\
P\bigg(\lim_{N\rightarrow\infty}\frac{N_1(V^{(i)}_{\ket{\psi_k}},N_k(\{\Pi_{l}\}_{\ket{\psi_i}},N_i(X,N)))}{N_k(\{\Pi_{l}\}_{\ket{\psi_i}},N_i(X,N))}\nonumber\\
=|\inpr{\psi_i}{\psi_k}|^2\bigg)=1.~~~~~~
\label{indvidualrelfrelln}
\end{eqnarray}
Using Eqs. (\ref{Fexptapp}), (\ref{indvidualrelfrelln}),  we obtain Eq. (\ref{LLN}).

\bibliographystyle{apsrev4-1}
\bibliography{bib_ch}

\end{document}